\documentclass[10pt]{article}

\usepackage[utf8]{inputenc}
\usepackage[T1]{fontenc}
\usepackage[top=0.85in, bottom=0.85in, left=1.0in, right=1.0in,
            headsep=0.3in, footskip=0.3in]{geometry}

\usepackage{fancyhdr}
\usepackage{xcolor}  

\usepackage{titlesec}
\titleformat{\section}
  {\normalsize\scshape\centering}
  {}{0pt}{}[\vspace{2pt}\hrule height 0.4pt\vspace{4pt}]
\titlespacing*{\section}{0pt}{1.2em}{0.6em}

\titleformat{\subsection}
  {\normalsize\itshape}
  {}{0pt}{}
\titlespacing*{\subsection}{0pt}{0.9em}{0.3em}

\usepackage{abstract}

\usepackage{titling}
\pretitle{%
  \begin{center}
    \rule{\linewidth}{1pt}\\[0.5em]
    \large\scshape
}
\posttitle{%
  \\[0.4em]\rule{\linewidth}{0.5pt}
  \end{center}
  \vspace{-1.0em}
}
\preauthor{\begin{center}\small\itshape}
\postauthor{\end{center}\vspace{-0.6em}}
\predate{}
\postdate{}

\usepackage{hyperref}
\hypersetup{
  colorlinks=false,
  hidelinks,
  pdftitle={107. Warfare},
  pdfauthor={Nathan Reitinger}
}
\usepackage[numbers,sort&compress]{natbib}
\usepackage{setspace}
\usepackage{multicol}

\usepackage{tcolorbox}
\tcbuselibrary{skins}
\newtcolorbox{sourcebox}{
  enhanced,
  colback=boxbg,
  colframe=boxrule,
  arc=3pt,
  boxrule=0.6pt,
  left=8pt, right=8pt, top=5pt, bottom=5pt,
  fontupper=\small,
}
\usepackage{xcolor}
\definecolor{ruleblue}{RGB}{70, 130, 180}
\definecolor{linkblue}{RGB}{30, 100, 200}
\definecolor{boxbg}{RGB}{240, 246, 255}
\definecolor{boxrule}{RGB}{100, 149, 237}

\begin{document}

\title{107.~Warfare}
\author{}
\date{}
\maketitle
\thispagestyle{fancy}
\fancyfoot[C]{\ifnum\value{page}=1\relax\else\footnotesize\thepage\fi}

\begin{abstract}
\noindent
What happens when artificial intelligence is given the power to make life-and-death
decisions? Is it ever desirable for a machine to decide if a human lives or dies, and
if so, what types of restrictions should be placed around weapons capable of making
these decisions? In the international law context, these types of weapons are known
as ``lethal autonomous weapons systems''---weapons that, after being turned on, select
and engage targets without human intervention. These weapons represent one of the many
difficult areas that have been upended by artificial intelligence. Accountability gaps,
alignment with ``just'' war practices, and moral or ethical qualms with machines taking
human lives are just a few of the headline issues in this area. This entry will provide
an overview of lethal autonomous weapons systems and break down the discussion along
technical and legal lines.

\medskip



\end{abstract}

\begin{sourcebox}
  \textbf{\textsc{Definition:}} 
  Lethal Autonomous Weapons Systems (LAWS) are weapons systems that select and engage targets without human intervention post-activation. In other words, these are weapons systems that may act independently when making kill–no–kill decisions.
\end{sourcebox}

\section{Introduction}

One of the largest use cases for artificial intelligence concerns the battlefield. Unsurprisingly, this particular use of artificial intelligence has been highly controversial. Several novel issues arise when machines—instead of humans—make life-and-death decisions: accountability gaps, morality in decision-making, and the future of war if it is automated. In this entry, the concept of lethal autonomous weapons systems (LAWS) is introduced and broken down into two sub-topics: technical considerations and legal considerations. The entry ends with an outlook for the future of LAWS.

\section{Technical}

In the autonomous weapons hierarchy, LAWS sit second from the top. At the bottom are in-the-loop systems. These are systems that require a human to ``complete the loop.'' These weapons systems do not work without human engagement (\textit{e.g.}, a single-action trigger pistol). Next up are on-the-loop systems. Here, the system may operate in a semi-autonomous mode, but a human remains a crucial component and is always able to intervene, though the machine is given some ability to make decisions on its own (\textit{e.g.}, placing a moving vehicle on ``cruise control'' automates the acceleration capabilities of a vehicle, though a human can take back control when necessary; likewise a self-aiming rifle may decide when to shoot, but a human selects the target and the weapon will not operate without that selection). Finally, out-of-the-loop systems are those that, once activated, select and engage targets without human intervention or control (\textit{i.e.}, LAWS).

\medskip

\noindent It is worth noting that even the definition of LAWS has seen a large amount of contention. For example, what is the difference between on-the-loop (\textit{i.e.}, also called semi-autonomous) and out-of-the-loop (\textit{i.e.}, also called fully autonomous) systems? The short answer is that, generally, the difference between second-loop and third-loop systems is one of degree, rather than kind. Consider that a commander may activate a sentry gun by turning it on (\textit{i.e.}, select and engage any moving object within a geographic location), but the commander could also turn off the sentry gun. Does this on-off ability make the sentry gun a second-loop system? This question reveals why the mostly accepted definition of LAWS is, in fact, narrow. LAWS refer specifically to situations where a weapons system both ``selects'' and ``engages'' targets. The sentry gun, once activated, fits this bill. The sentry gun will select and engage targets on its own; therefore, this is an example of a third-loop system. It is also noteworthy that, practically, out-of-the-loop systems may permit human oversight, but are often designed to act in situations where human oversight is impossible or impractical, as is the case with the sentry gun, the United States Navy's Phalanx system, or Israel's Iron Dome. Each of these weapons systems may allow for human control, but a human's ability to monitor and react pales in comparison to the machine.

\medskip

\noindent Some of the confusion in demarcating first- to second- to third-loop systems may come from thinking that current-day technology is more capable than it really is. If there existed a machine that not only selects and engages targets on its own, but also possesses something akin to moral agency (\textit{e.g.}, debating the merits of actions with logical reasoning or seeming to express free will), then the definition of ``fully'' autonomous seems like it should take on a new meaning besides selection and engagement. This type of autonomous system, ``a loop of its own,'' muddies the waters because it points toward a murky area where a human is truly being replaced by a machine—a true accountability gap: If a machine, on its own accord, committed an untoward action, why should anyone but the machine be responsible for that action? To be clear, these questions are unnecessary at the current stage of development. Fourth-loop systems like those do not yet exist, and even the artificial general intelligence being developed right now may not qualify as anything but a fully autonomous weapon. Today, only tools wielded by operators exist, even if the orders given to these tools have changed from a required physical gesture to turning on a device that comes with selection and engagement functionality.

\medskip

\noindent A final technical consideration for LAWS relates to the development of these weapons, which happens to be ongoing and occurring in parallel with consumer products. For a weapons system to ``select'' and ``engage'' a target, it must have the ability to both locate that target and make a reasoned decision on whether to engage the target. The Iron Dome previously mentioned, for example, is able to ``see'' its surroundings and identify objects through the use of radar. This is the first step, selection. The system then, once an object is identified, must decide whether an object is a viable target (\textit{e.g.}, direction and speed of the object). This is the second step, engagement. These two properties, together, are also a development focus of civilian products being infused with artificial intelligence. And this is an area experiencing tremendous change. Autonomous vehicles, for instance, an unheard-of technology ten years ago, must: (1) see the road to know whether a vehicle is in its appropriate lane, and (2) make decisions based on that vision, like whether a stoplight is green, or what to do when a pedestrian is crossing the street. Two points are important here.

\medskip

\noindent First, the underlying technology, computer vision, in each use case, is very similar, though the training data for each model may differ. Therefore, it is worth noting that the consumer-focused companies developing artificial intelligence are, by improving consumer products with artificial intelligence, in some ways, pushing the cutting-edge for military developments as well. Second, making long-lasting decisions about a technology that is in a rapid state of development is difficult. Banning or regulating technologies that do not exist yet or may look very different five years from now is a difficult proposition.

\section{Legal}

Before discussing the legal context for LAWS, it is worth being explicit about the benefits and drawbacks of LAWS—in turn showing why it is not surprising that sovereign nations are having a difficult time forming a unified opinion on whether LAWS are a permissible form of weaponry. On the one hand, LAWS may be appealing from a military perspective. As Gordon Johnson of Joint Forces Command at the Pentagon stated in 2005: “They don’t get hungry. They’re not afraid. They don’t forget their orders. They don’t care if the guy next to them has just been shot. Will they do a better job than humans? Yes.” Additionally, LAWS may be more transparent about decisions than human soldiers, as research suggests that up to 45\% of soldiers and 60\% of Marines would not report a fellow soldier if that soldier injured or killed an innocent combatant. On the other hand, a host of issues are readily apparent: Computer systems like LAWS are routinely hacked and set to behave in undesirable ways, the systems powering LAWS come with a hefty number of biases, computers routinely make mistakes and experience sometimes-catastrophic software bugs, and there are moral and ethical issues, omitted here for brevity, with machines making life-and-death decisions in the first place.

\medskip

\noindent These conflicting perspectives, along with a host of geopolitical factors, have created a mixed bag when it comes to the legality of using LAWS. Many countries were quick to support a preemptive ban on LAWS prior to their use on the battlefield. Missing among proponents of a ban, however, were countries like Israel, Russia, the United Kingdom, the United States, and South Korea. That voluntary preemptive actions like outright bans have not reached unanimity, however, does not mean LAWS are always legal.

\medskip

\noindent The use of LAWS on the battlefield must nonetheless comply with international humanitarian law (IHL) and international criminal law. Most relevant here, IHL requires that all “legal” weapons systems maintain a series of principles, including necessity, distinction, and proportionality. In general terms, the machine could not be used unless it were able to understand context in terms of the appropriate use of force and be able to distinguish between appropriate and inappropriate targets (\textit{e.g.}, combatants versus civilians). Controversy exists over whether, technically, machines now, or in the future, will be able to satisfy those constraints.

\medskip

\noindent Likewise, there are concerns over the decisions made by LAWS given the potential for an accountability gap that could incentivize lawlessness—in turn casting doubt on any type of permitted use of LAWS. The argument, put simply, is that the machine's actions will lack a mens rea, thereby obviating the ability of criminal law to incentivize responsible use. At least in part, these concerns may be tempered when considering that only third-loop systems exist. A human operator will have nevertheless made a decision to use the autonomous weapons system in the first place, and that human operator, though somewhat removed from what could be an untoward action, may be a better candidate for culpability than the tool being used. On the other hand, the liability-gap argument is most effective when considering that machines may make unpredictable or undesired actions, and those actions may seem unfairly pinned on an original decision-maker. In turn, discussions over accountability and compliance with IHL are ongoing and will likely continue to be ongoing due to rapid innovation, political tensions, and issues over international harmonization.

\section{Conclusion}

Automated weapons systems are currently and will continue to be found on the battlefield. These systems are increasingly capable of replacing human soldiers, could potentially change the way war operates on a global scale, and present complex ethical, moral, and legal issues that, like the doctrine of \textit{jus cogens}, are difficult to reach a global consensus on. Countries around the world are continuing to support a ban on LAWS, but holdouts remain, adding support for the regulation of LAWS in a way that creates reasonable boundaries on use rather than preemptive bans.



\nocite{*}
{\small
\setlength{\bibsep}{2pt}
\bibliography{warfare}
}

\end{document}